\documentclass[12pt]{article}
\usepackage{amsmath,amssymb}
\usepackage{textcomp}
\usepackage{array,multirow,graphicx}
\DeclareMathAlphabet{\pazocal}{OMS}{zplm}{m}{n}
\usepackage{graphicx}
\usepackage{caption}
\usepackage{url}
\usepackage{subcaption}
\usepackage{authblk}
\usepackage[margin=0.59in]{geometry}
\usepackage[square,numbers]{natbib}
\date { }

\title{End-to-End Speech-Driven Facial Animation with Temporal GANs}

\newcommand\figref{Fig.~\ref}
\newcommand\tabref{Table~\ref}
\newcommand\eref{Equation~\ref}
\def\b{\textbf}
\DeclareMathOperator{\EX}{\mathbb{E}}% expected value

\author[1]{Konstantinos Vougioukas}
\author[1,2]{Stavros Petridis}
\author[1,2]{Maja Pantic}
\affil[1]{iBUG Group, Imperial College London}
\affil[2]{Samsung AI Centre, Cambridge, UK}

\begin{document}

\maketitle

\begin{abstract}
Speech-driven facial animation is the process which uses speech signals to automatically synthesize a talking character. The majority of work in this domain creates a mapping from audio features to visual features. This often requires post-processing using computer graphics techniques to produce realistic albeit subject dependent results. We present a system for generating videos of a talking head, using a still image of a person and an audio clip containing speech, that doesn't rely on any handcrafted intermediate features. To the best of our knowledge, this is the first method capable of generating subject independent realistic videos directly from raw audio. Our method can generate videos which have (a) lip movements that are in sync with the audio and (b) natural facial expressions such as blinks and eyebrow movements \footnote{Videos are  available here: \normalsize{\textbf{\url{https://sites.google.com/view/facialsynthesis/home} }}}. We achieve this by using a temporal GAN with 2 discriminators, which are capable of capturing different aspects of the video. The effect of each component in our system is quantified through an ablation study. The generated videos are evaluated based on their sharpness, reconstruction quality, and lip-reading accuracy. Finally, a user study is conducted, confirming that temporal GANs lead to more natural sequences than a static GAN-based approach.
\end{abstract}

%------------------------------------------------------------------------- 
\begin{figure}[h]
\begin{center}
\includegraphics[width=0.99\textwidth]{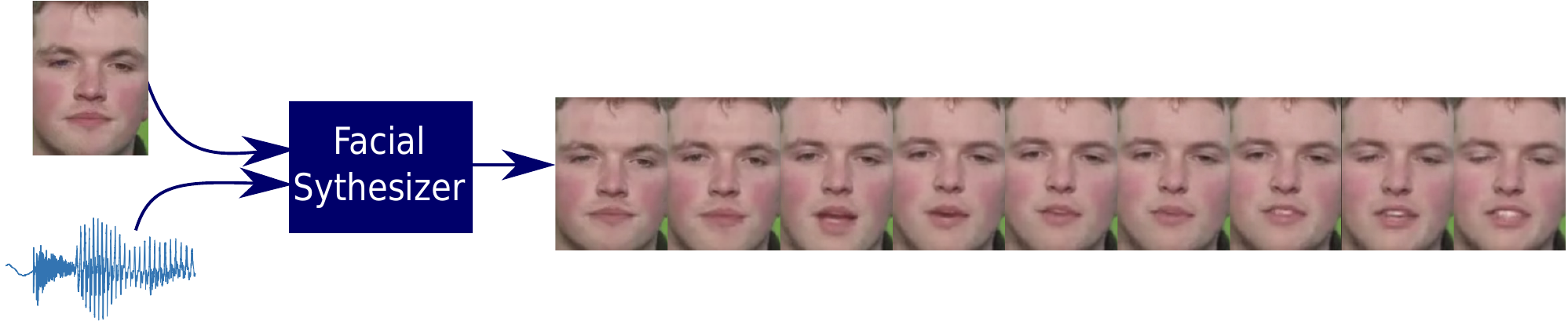}
\end{center}
\caption{The proposed end-to-end face synthesis model, capable of producing realistic sequences of faces using one still image and an audio track containing speech. The generated sequences exhibit smoothness and natural expressions such as blinks and frowns.}
\label{fig:summary}
\end{figure}

\section{Introduction}
\label{sec:intro}

Facial animation plays a major role in computer generated imagery because the face is the primary outlet of information. The problem of generating realistic talking heads is multifaceted, requiring high-quality faces, lip movements synchronized with the audio, and plausible facial expressions. This is especially challenging because humans are adept at picking up subtle abnormalities in facial motion and audio-visual synchronization.

Of particular interest is speech-driven facial animation since speech acoustics are highly correlated with facial movements \cite{Yehia1998QuantitativeBehavior}. These systems could simplify the film animation process through automatic generation from the voice acting. They can also be applied in movie dubbing to achieve better lip-syncing results. Moreover, they can be used to generate parts of the face that are occluded or missing in a scene. Finally, this technology can improve band-limited visual telecommunications by either generating the entire visual content based on the audio or filling in dropped frames.

The majority of research in this domain has focused on mapping audio features (e.g. MFCCs) to visual features (e.g. landmarks, visemes) and using computer graphics (CG) methods to generate realistic faces \cite{Karras2017}. Some methods avoid the use of CG by selecting frames from a person-specific database and combining them to form a video \cite{Bregler1997,Suwajanakorn2017}. Regardless of which approach is used these methods are subject dependent and are often associated with a considerable overhead when transferring to new speakers.

Subject independent approaches have been proposed that transform audio features to video frames \cite{Chung2017} but there is still no method to directly transform raw audio to video. Furthermore, many methods restrict the problem to generating only the mouth. Even techniques that generate the entire face are primarily focused on obtaining realistic lip movements, and typically neglect the importance of generating natural facial expressions.

Some methods generate frames based solely on present information \cite{Chung2017}, without taking into account the facial dynamics. This makes generating natural sequences, which are characterized by a seamless transition between frames, challenging. Some video generation methods have dealt with this problem by generating the entire sequence at once \cite{Vondrick2016} or in small batches \cite{Saito2016}. However, this introduces a lag in the generation process, prohibiting their use in real-time applications and requiring fixed length sequences for training.

We propose a temporal generative adversarial network (GAN), capable of generating a video of a talking head from an audio signal and a single still image (see \figref{fig:summary} ). First, our model captures the dynamics of the entire face producing not only synchronized mouth movements but also natural facial expressions, such as eyebrow raises, frowns and blinks. This is due to the use of an RNN-based generator and sequence discriminator, which also gives us the added advantage of handling variable length sequences. Natural facial expressions play a crucial role in achieving truly realistic results and their absence is often a clear tell-tale sign of generated videos. This is exploited by methods such as the one proposed in \cite{LiInBlinking}, which detects synthesized videos based on the existence of blinks.

Secondly, our method is subject independent, does not rely on handcrafted audio or visual features, and requires no post-processing. To the best of our knowledge, this is the first end-to-end technique that generates talking faces directly from the raw audio waveform.

The third contribution of this paper is a comprehensive assessment of the performance of the proposed method. An ablation study is performed on the model in order to quantify the effect of each component in the system. We measure the image quality using popular reconstruction and sharpness metrics, and compare it to a non-temporal approach. Additionally, we propose using lip reading techniques to verify the accuracy of the spoken words and face verification to ensure that the identity of the speaker is maintained throughout the sequence. Evaluation is performed in a subject independent way on the GRID \cite{Cooke2006} and TCD TIMIT \cite{Harte2015} datasets, where our model achieves truly natural results. Finally, the realism of the videos is assessed through an online Turing test, where users are shown videos and asked to identify them as either real or generated.

\section{Related Work}
\label{sec:related}
\subsection{Speech-Driven Facial Animation}
\label{sec:speech_driven_animation}
The problem of speech-driven video synthesis is not new in computer vision and has been the subject of interest for decades. Yehia {\em et al.} \cite{Yehia1998QuantitativeBehavior} first examined the relationship between acoustics, vocal-tract and facial motion, showing a strong correlation between visual and audio features and a weak coupling between head motion and the fundamental frequency of the speech signal \cite{Yehia2002LinkingAcoustics}.

Some of the earliest methods for facial animation relied on hidden Markov models (HMMs) to capture the dynamics of the video and speech sequences. Simons and Cox \cite{Simons1990} used vector quantization to achieve a compact representation of video and audio features, which were used as the states for their HMM. The HMM was used to recover the most likely mouth shapes for a speech signal. A similar approach is used in \cite{Yamamoto1998} to estimate the sequence of lip parameters. Finally, the {\em Video Rewrite} \cite{Bregler1997} method relies on the same principals to obtain a sequence of triphones that are used to look up mouth images from a database.

Although HMMs were initially preferred to neural networks due to their explicit breakdown of speech into intuitive states, recent advances in deep learning have resulted in neural networks being used in most modern approaches. Like past attempts, most of these methods aim at performing a feature-to-feature translation. A typical example of this is \cite{Taylor2017}, which uses a deep neural network (DNN) to transform a phoneme sequence into a sequence of shapes for the lower half of the face. Using phonemes instead of raw audio ensures that the method is subject independent. Similar deep architectures based on recurrent neural networks (RNNs) have been proposed in \cite{Fan2015,Suwajanakorn2017}, producing realistic results but are subject dependent and require retraining or re-targeting steps to adapt to new faces. 

Convolutional neural networks (CNN) are used in \cite{Karras2017} to transform audio features to 3D meshes of a specific person. This system is conceptually broken into sub-networks responsible for capturing articulation dynamics and estimating the 3D points of the mesh. Finally, Chung {\em et al.} \cite{Chung2017} proposed a CNN applied on Mel-frequency cepstral coefficients (MFCCs) that generates subject independent videos from an audio clip and a still frame. The method uses an $L_1$ loss at the pixel level resulting in blurry frames, which is why a deblurring step is also required. Secondly, this loss at the pixel level penalizes any deviation from the target video during training, providing no incentive for the model to produce spontaneous expressions and resulting in faces that are mostly static except for the mouth. 

\subsection{GAN-Based Video Synthesis}
\label{sec:gans}
The recent introduction of GANs in \cite{Goodfellow2014} has shifted the focus of the machine learning community to generative modelling. GANs consist of two competing networks: a generative network and a discriminative network. The generator's goal is to produce realistic samples and the discriminator's goal is to distinguish between the real and generated samples. This competition eventually drives the generator to produce highly realistic samples. GANs are typically associated with image generation since the adversarial loss produces sharper, more detailed images compared to $L_1$ and  $L_2$ losses. However, GANs are not limited to these applications and can be extended to handle videos \cite{Mathieu2015,Li2017,Vondrick2016,Tulyakov2017}. 

Straight-forward adaptations of GANs for videos are proposed in \cite{Vondrick2016, Saito2016}, replacing the 2D convolutional layers with 3D convolutional layers. Using 3D convolutions in the generator and discriminator networks is able to capture temporal dependencies but requires fixed length videos. This limitation was overcome in \cite{Saito2016} but constraints need to be imposed in the latent space to generate consistent videos.

The {\em MoCoGAN} system proposed in \cite{Tulyakov2017} uses an RNN-based generator, with separate latent spaces for motion and content. This relies on the empirical evidence shown in \cite{Radford2015} that GANs perform better when the latent space is disentangled. {\em MoCoGAN} uses a 2D and 3D CNN discriminator to judge frames and sequences respectively. A sliding window approach is used so that the 3D CNN discriminator can handle variable length sequences.

GANs have also been used in a variety of cross-modal applications, including text-to-video and audio-to-video. The text-to-video model proposed in \cite{Li2017} uses a combination of variational auto encoders (VAE) and GANs in its generating network and a 3D CNN as a sequence discriminator. Finally, Chen {\em et al.} \cite{Chen1998} propose a GAN-based encoder-decoder architecture that uses CNNs in order to convert audio spectrograms to frames and vice versa.

\section{End-to-End Speech-Driven Facial Synthesis}
\label{sec:model}
\begin{figure}[h]
\begin{center}
\includegraphics[width=0.99\textwidth]{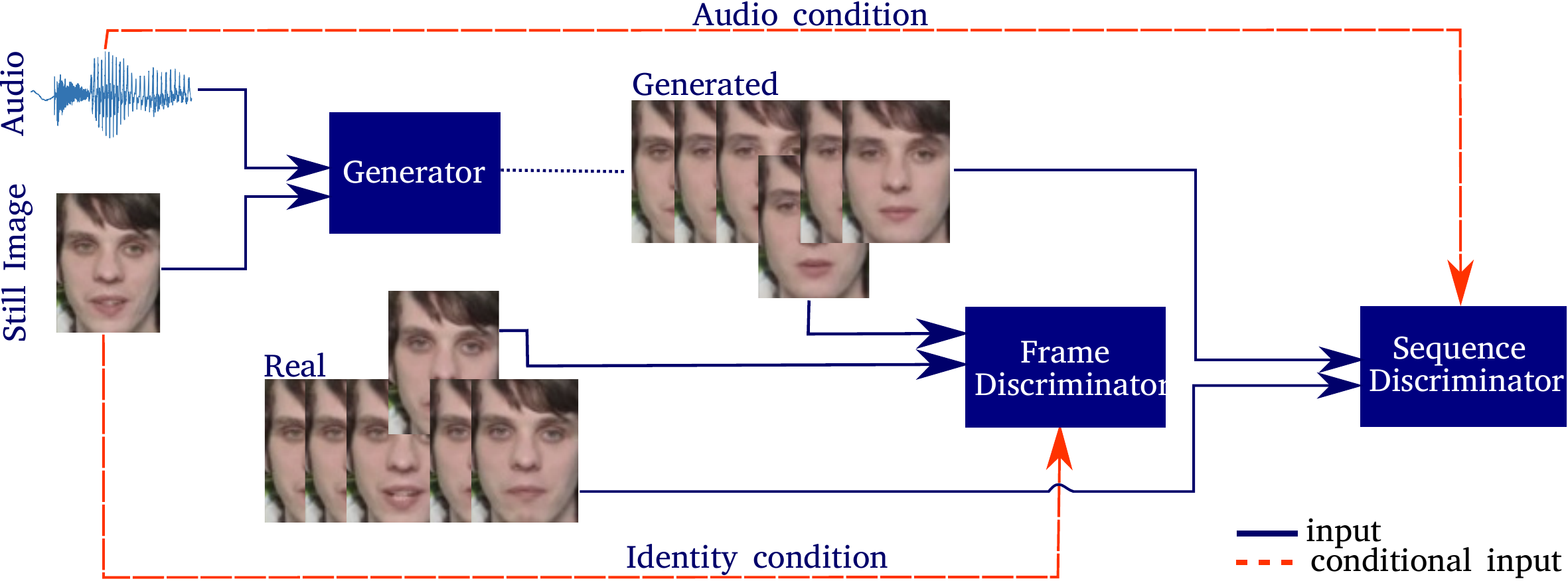}
\end{center}
\caption{The deep model for speech-driven facial synthesis. This uses 2 discriminators to incorporate the different aspects of a realistic video. Details about the architecture are presented in the supplementary material.}
\label{fig:model}
\end{figure}

The proposed architecture for speech-driven facial synthesis is shown in \figref{fig:model}. The system is made up of a generator and 2 discriminators, each of which evaluates the generated sequence from a different perspective. The capability of the generator to capture various aspects of natural sequences is directly proportional to the ability of each discriminator to discern videos based on them.

\subsection{Generator}
\label{sec:generator}

The inputs to the generator networks consist of a single image and an audio signal, which is divided into overlapping frames each corresponding to $0.16$ seconds. The generator network in this architecture can be conceptually divided into subnetworks as shown in \figref{fig:gendisc}. Using an RNN-based generator allows us to synthesize videos frame-by-frame, which is necessary for real-time applications.

\subsubsection{Identity Encoder}
\label{sec:id_encoder}

The speaker's identity is encoded using a 6 layer CNN. Each layer uses strided 2D convolutions, followed by batch normalization and ReLU activation functions. The {\em Identity Encoder} network reduces the input image to a $50$ dimensional encoding $z_{id}$.

\subsubsection{Context Encoder}
\label{sec:context_encoder}
Audio frames are encoded using a network comprising of 1D convolutions followed by batch normalization and ReLU activations. The initial convolutional layer starts with a large kernel, as recommended in \cite{Dai2016}, which helps limit the depth of the network while ensuring that the low-level features are meaningful. Subsequent layers use smaller kernels until an embedding of the desired size is achieved. The audio frame encodings are input into a 2 layer GRU, which produces a context encoding $z_c$ with $256$ elements.

\subsubsection{Frame Decoder}
\label{sec:frame_decoder}
The identity encoding $z_{id}$ is concatenated to the context encoding $z_c$ and a noise component $z_n$ to form the latent representation. The 10-dimensional $z_n$ vector is obtained from a {\em Noise Generator}, which is a 1-layer GRU that takes Gaussian noise as input. The {\em Frame Decoder} is a CNN that uses strided transposed convolutions to produce the video frames from the latent representation. A U-Net \cite{Ronneberger2015} architecture is used with skip connections between the {\em Identity Encoder} and the {\em Frame Decoder} to help preserve the identity of the subject.

\begin{figure}[ht]
  \centering
  \begin{subfigure}[b]{0.5\linewidth}
    \centering\includegraphics[width=0.97\textwidth]{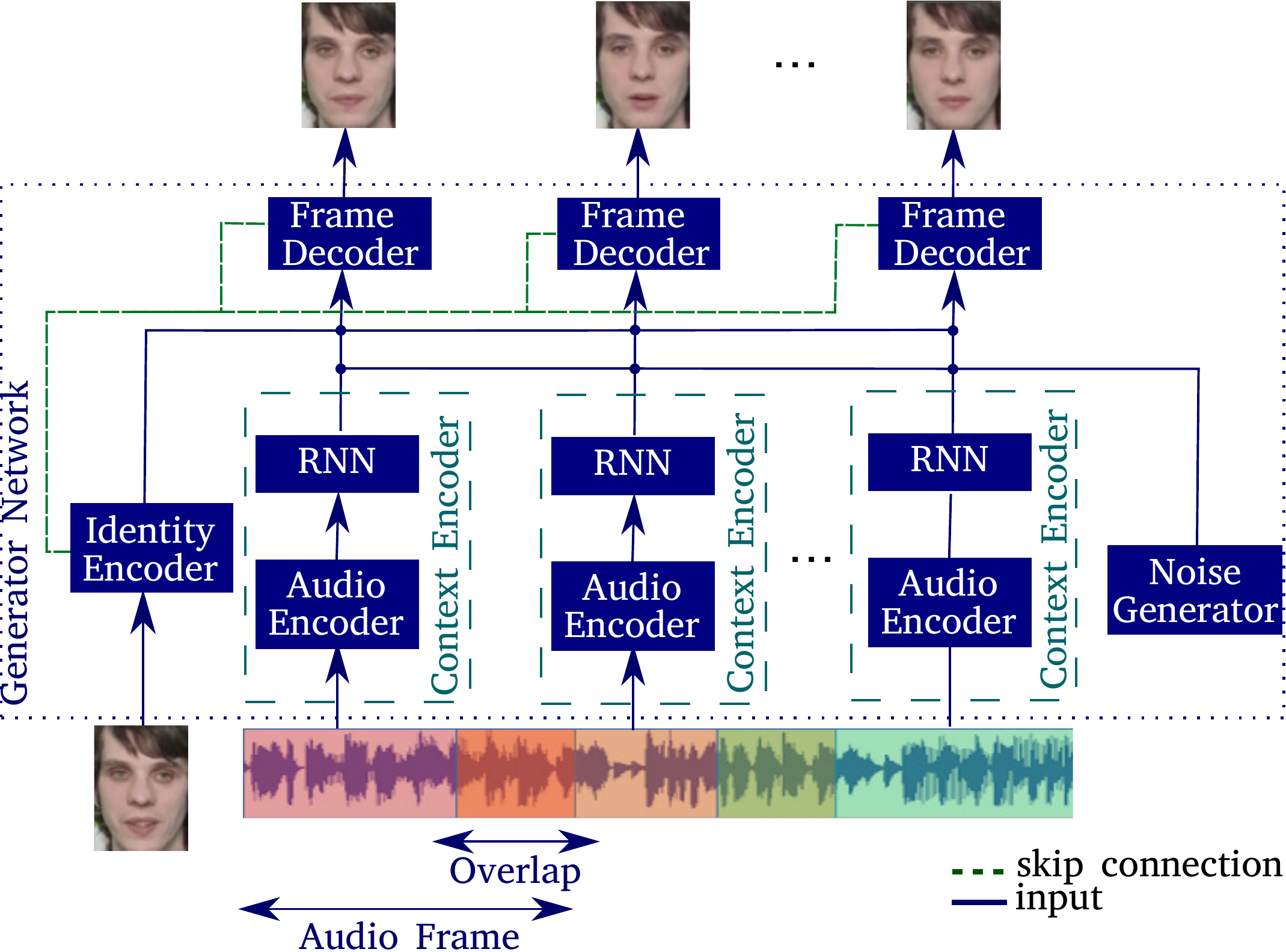}
    \caption{\label{fig:gendisc_gen} Generator}
  \end{subfigure}%
  \begin{subfigure}[b]{0.5\linewidth}
    \centering\includegraphics[width=0.97\textwidth]{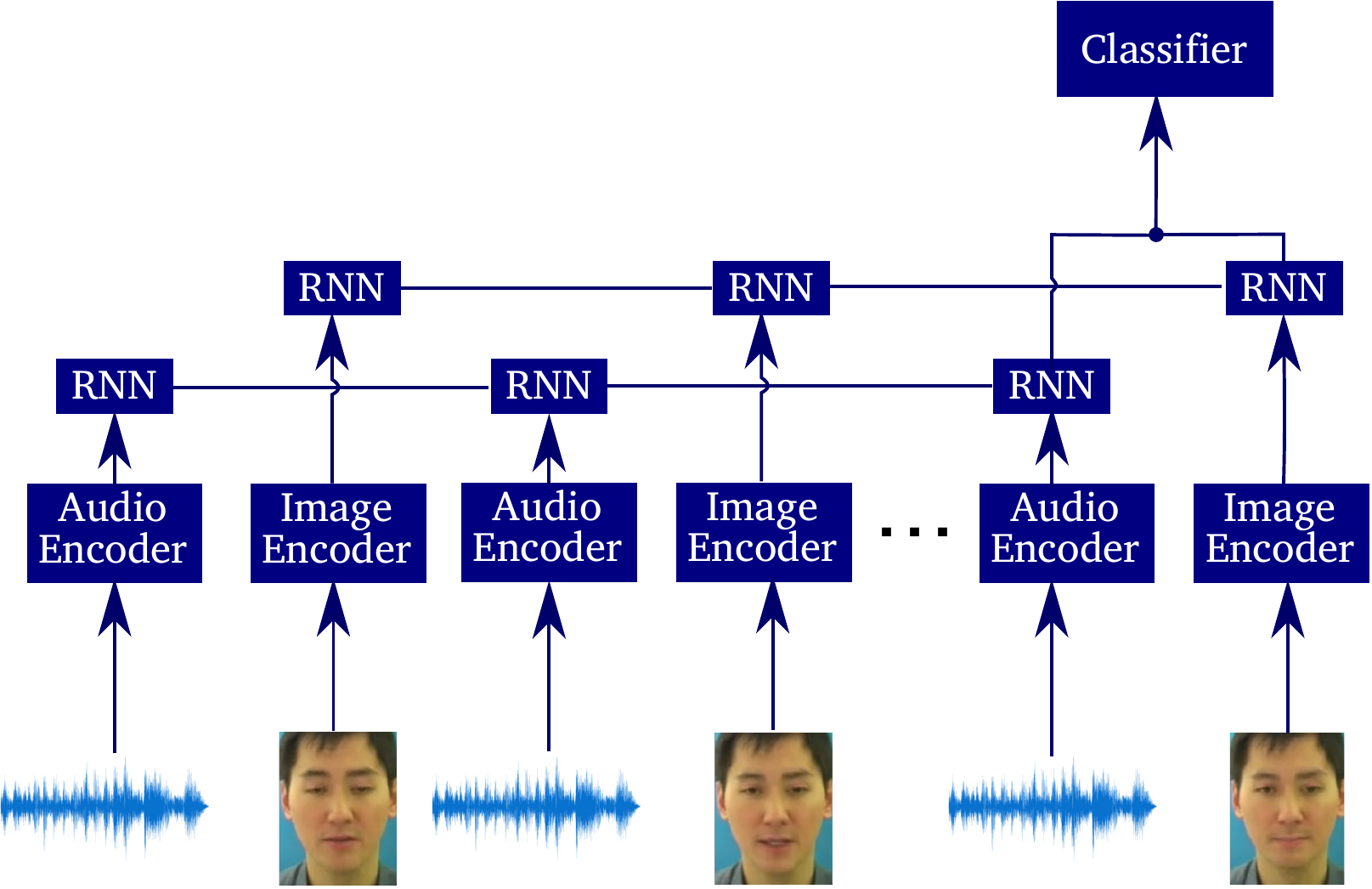}
    \caption{\label{fig:gendisc_disc} Sequence Discriminator}
  \end{subfigure}
  \caption{The architecture of the (a) Generator which consists of a {\em Context Encoder} (audio encoder and RNN), an {\em Identity Encoder}, a {\em Frame Decoder} and {\em Noise Generator} (b) Sequence Discriminator, consisting of an audio encoder, an image encoder, GRUs and a small classifier.}
\label{fig:gendisc}
\end{figure}

\subsection{Discriminators}
\label{sec:discriminators}
Our system has two different types of discriminator. The {\em Frame Discriminator} helps achieve a high-quality reconstruction of the speakers' face throughout the video. The {\em Sequence Discriminator} ensures that the frames form a cohesive video which exhibits natural movements and is synchronized with the audio. 

\subsubsection{Frame Discriminator}
\label{sec:frame_disc}
The {\em Frame Discriminator} is a 6-layer CNN that determines whether a frame is real or not. Adversarial training with this discriminator ensures that the generated frames are realistic. The original still frame is used as a condition in this network, concatenated channel-wise to the target frame to form the input as shown in \figref{fig:gendisc}. This enforces the person's identity on the frame. 

\subsubsection{Sequence Discriminator}
\label{sec:seq_disc}
The {\em Sequence Discriminator} presented in \figref{fig:gendisc} distinguishes between real and synthetic videos. The discriminator receives a frame at every time step, which is encoded using a CNN and then fed into a 2-layer GRU. A small (2-layer) classifier is used at the end of the sequence to determine if the sequence is real or not. The audio is added as a conditional input to the network, allowing this discriminator to classify speech-video pairs.

\subsection{Training}
\label{sec:training}
The {\em Frame discriminator} ($D_{img}$) is trained on frames that are sampled uniformly from a video $x$ using a sampling function $S(x)$. The {\em Sequence discriminator} ($D_{seq}$), classifies based on the entire sequence $x$ and audio $a$. The loss of our GAN is an aggregate of the losses associated with each discriminator as shown in \eref{eq:adv_loss}.
\begin{equation}
\begin{split}
\pazocal{L}_{adv}(D_{img}, D_{Seq}, G) = & \EX_{x \sim P_d}[\log D_{img}(S(x), x_1)] + \EX_{z \sim P_z}[\log (1- D_{img}(S(G(z)), x_1))] + \\
& \EX_{x \sim P_d}[\log D_{seq}(x, a)] + \EX_{z \sim P_z}[\log (1- D_{seq}(G(z), a))]
\end{split}
\label{eq:adv_loss}
\end{equation}

An $L_1$ reconstruction loss is also used to improve the synchronization of the mouth movements. However we only apply the reconstruction loss to the lower half of the image since it discourages the generation of facial expressions. For a ground truth frame $F$ and a generated frame $G$ with dimensions $W \times H$ the reconstruction loss at the pixel level is:
\begin{equation}
\pazocal{L}_{L_1} = \sum_{p \in [ 0, W] \times [ \frac{H}{2}, H ] }|F_{p} - G_{p}|
\label{eq:reconstruction_loss}
\end{equation}

The final objective is to obtain the optimal generator $G^*$, which satisfies \eref{eq:loss}. The model is trained until no improvement is observed on the reconstruction metrics on the validation set for 10 epochs. The $\lambda$ hyperparameter controls the contribution of each loss factor and was set to $400$ following a tuning procedure on the validation set.
\begin{equation}
\begin{split}
\arg \min_{G} \max_{D}\pazocal{L}_{adv} + \lambda \pazocal{L}_{L_1}
\end{split}
\label{eq:loss}
\end{equation}

We used Adam \cite{Kingma2014} for all the networks with a learning rate of $0.0002$ for the {\em Generator} and $0.001$ {\em Frame Discriminator} which decay after epoch 20 with a rate of $10\%$. The {\em Sequence Discriminator} uses a smaller fixed learning rate of $5\cdot 10^{-5}$.

\section{Experiments}
\label{sec:experiments}
Our model is implemented in PyTorch  and takes approximately a week to train using an Nvidia GeForce GTX 1080 Ti GPU. During inference the average generation time per frame is 7ms on the GPU, permitting the use of our method use in real time applications. A sequence of 75 frames can be synthesized in 0.5s. The frame and sequence generation times increase to 1s  and 15s respectively when processing is done on the CPU.

\subsection{Datasets}
\label{sec:datasets}
The GRID dataset has 33 speakers each uttering 1000 short phrases, containing 6 words taken from a limited dictionary. The TCD TIMIT dataset has 59 speakers uttering approximately 100 phonetically rich sentences each. We use the recommended data split for the TCD TIMIT dataset but exclude some of the test speakers and use them as a validation set. For the GRID dataset speakers are divided into training, validation and test sets with a $50\% - 20\% - 30\%$ split respectively. As part of our preprocessing all faces are aligned to the canonical face and images are normalized. We increase the size of the training set by mirroring the training videos. The amount of data used for training and testing is presented in \tabref{tab:Subjects}.

\begin{table}[h!]
\begin{center}
\begin{tabular}{|l|c|c|c|c|c|c|}
\hline
Dataset & Samples (Tr) & Hours (Tr) & Samples (V)& Hours (V)&  Samples (T)& Hours (T) \\
\hline\hline
GRID   & 31639 & 26.36 & 6999 & 5.83 & 9976 & 8.31 \\
TCD    & 8218  & 9.13 & 686 & 0.76 &977 & 1.09 \\
\hline
\end{tabular}
\end{center}
\caption{The samples and hours of video in the training (Tr), validation (V) and test (T) sets.}
\label{tab:Subjects}
\end{table}
\subsection{Metrics}
\label{sec:metrics}

We use common reconstruction metrics such as the peak signal-to-noise ratio (PSNR) and the structural similarity (SSIM) index to evaluate the generated videos. During the evaluation it is important to take into account the fact that reconstruction metrics penalize videos for any spontaneous expression. The frame sharpness is evaluated using the cumulative probability blur detection (CPBD) measure \cite{Narvekar2009}, which determines blur based on the presence of edges in the image and the frequency domain blurriness measure (FDBM) proposed in \cite{De2013},  which is based on the spectrum of the image. For these metrics larger values imply better quality.

The content of the videos is evaluated based on how well the video captures identity of the target and on the accuracy of the spoken words. We verify the identity of the speaker using the average content distance (ACD) \cite{Tulyakov2017}, which measures the average Euclidean distance of the still image representation, obtained using OpenFace \cite{Amos2016}, from the representation of the generated frames. The accuracy of the spoken message is measured using the word error rate (WER) achieved by a pre-trained lip-reading model. We use the LipNet model \cite{Assael2016}, which surpasses the performance of human lipreaders on the GRID dataset. For both content metrics lower values indicate better accuracy.

\subsection{Ablation Study}
\label{sec:ablation}
\begin{figure}[h!]
  \centering
  \begin{subfigure}[b]{0.5\linewidth}
    \centering\includegraphics[width=0.99\textwidth]{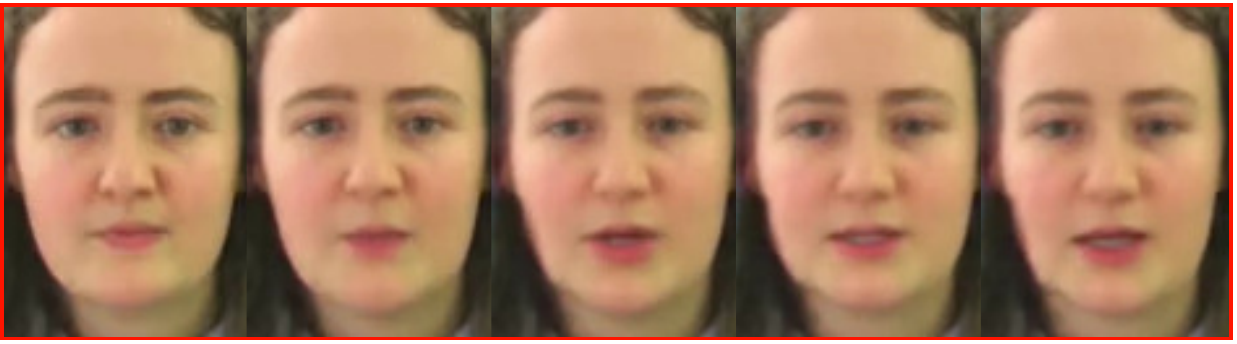}
    \caption{\label{fig:bluriness_l1} $L_1$ loss on entire frame}
  \end{subfigure}%
  \begin{subfigure}[b]{0.5\linewidth}
    \centering\includegraphics[width=0.99\textwidth]{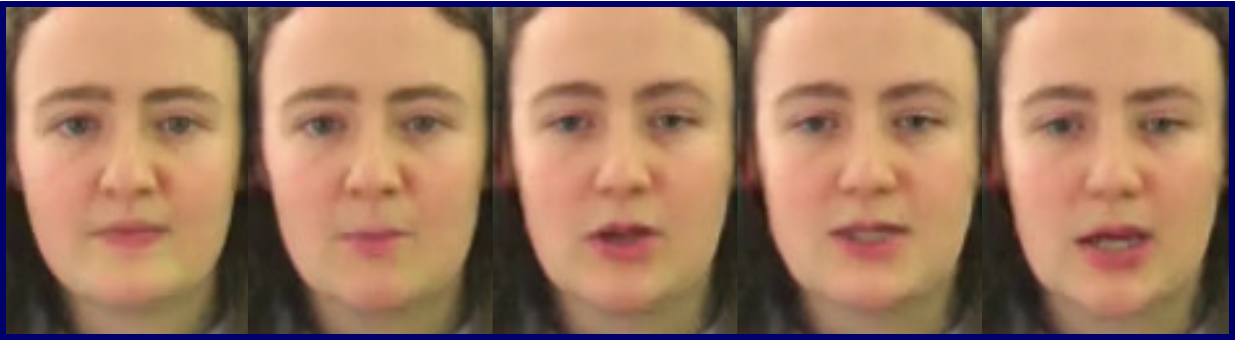}
    \caption{\label{fig:bluriness_model}  Proposed loss on frames}
  \end{subfigure}
  \caption{Frames using (a) only an $L_1$ loss on the entire face compared to (b) frames produced using the proposed method. Frames generated using an $L_1$ loss on the entire face (a) are blurrier than those produced from the proposed method (b).}
\label{fig:bluriness}
\end{figure}

In order to quantify the effect of each component of our system we perform an ablation study on the GRID dataset (see \tabref{tab:ablation}). We use the metrics from section \ref{sec:metrics} and a pre-trained LipNet model which achieves a WER of $21.4 \%$ on the ground truth videos. The average value of the ACD for ground truth videos of the same person is $0.74 \cdot 10^{-4}$ whereas for different speakers it is $1.4 \cdot 10^{-3}$. The $L_1$ loss achieves slightly better PSNR and SSIM results, which is expected as it does not generate spontaneous expressions, which are penalized by these metrics unless they happen to coincide with those in ground truth videos. This variation introduced when generating expressions is likely the reason for the small increase in ACD. The blurriness is minimized when using the adversarial loss as indicated by the higher FDBM and CPBD scores and \figref{fig:bluriness}. Finally, the effect of the sequence discriminator is shown in the lip-reading result achieving a low WER.

\begin{table}[h!]
\begin{center}
\begin{tabular}{|l|c|c|c|c|c|c|}
\hline
\multicolumn{1}{|l|}{Method} & \multicolumn{1}{c|}{PSNR} & \multicolumn{1}{c|}{SSIM} & \multicolumn{1}{c|}{FDBM} & \multicolumn{1}{c|}{CPBD} & \multicolumn{1}{c|}{ACD} & \multicolumn{1}{c|}{WER}\\
\hline\hline
Ground Truth Videos & N/A & N/A & 0.121 & 0.281 & 0.74 $\cdot 10^{-4}$ & 21.40\% \\
$L_1$ loss              & \b{28.47} & \b{0.859} & 0.101     & 0.210      & \b{	0.90 $\cdot 10^{-4}$} & 27.90\% \\
$L_1 + Adv_{img}$     & 27.71     & 0.840     & 0.114     & 0.274      & 1.04 $\cdot 10^{-4}$& 27.94\% \\
$L_1 + Adv_{img} + Adv_{seq}$     & 27.98     & 0.844     & \b{0.114} & \b{0.277}  & 1.02 $\cdot 10^{-4}$ & \textbf{25.45}\% \\
\hline
\end{tabular}
\end{center}
\caption{Assessment of each model in the ablation study performed on the GRID dataset}
\label{tab:ablation}
\end{table}

\subsection{Qualitative Results}
\label{sec:qualitative}
\begin{figure}[h!]
  \centering
  \begin{subfigure}[b]{0.49\linewidth}
    \centering\includegraphics[width=0.99\textwidth]{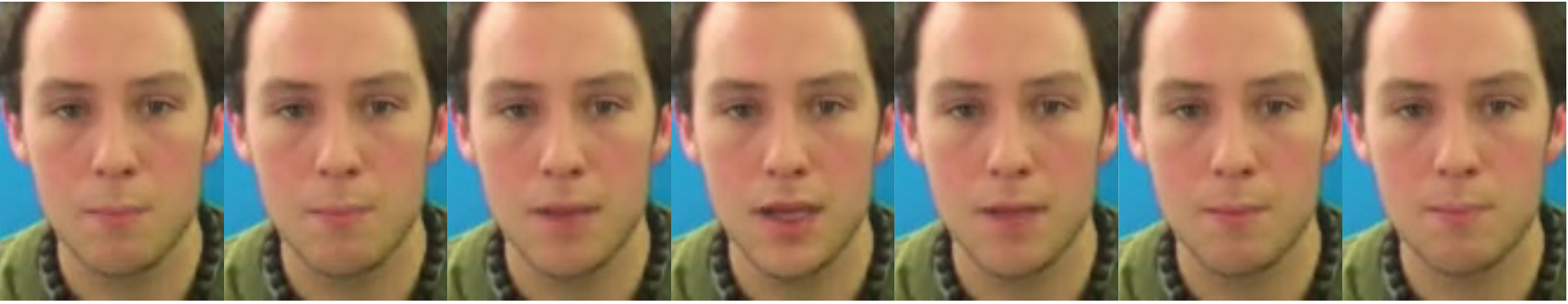}
    \caption{\label{fig:diff_audioa}  Female voice uttering the word ``bin''}
  \end{subfigure}
  \begin{subfigure}[b]{0.49\linewidth}
    \centering\includegraphics[width=0.99\textwidth]{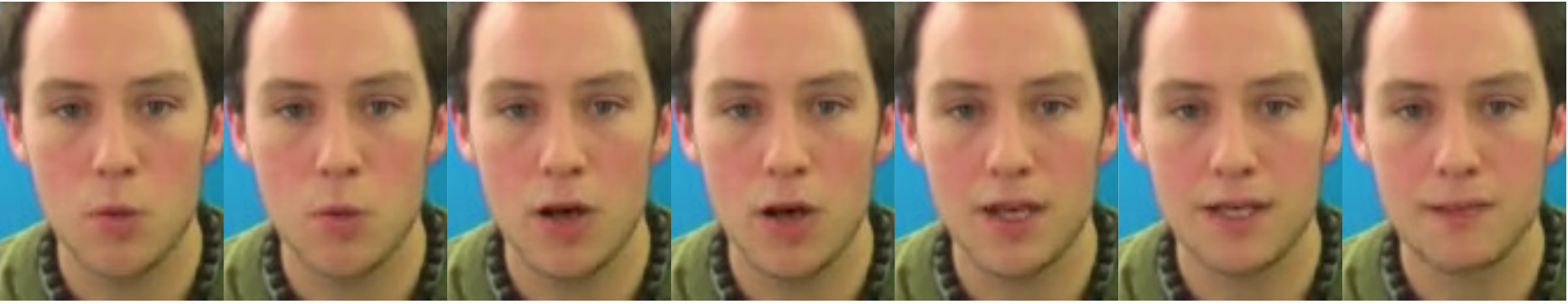}
    \caption{\label{fig:diff_audiob}   Male voice uttering the word ``white''}
  \end{subfigure}
  \caption{Generated sequences for (a) the word ``bin'' (b) the word ``white'' from the GRID test set. Coarticulation is evident in (a) where ``bin'' is followed by the word ``blue''.}
\label{fig:diff_audio}
\end{figure}

\begin{figure}[h!]
\begin{center}
\includegraphics[width=0.99\textwidth]{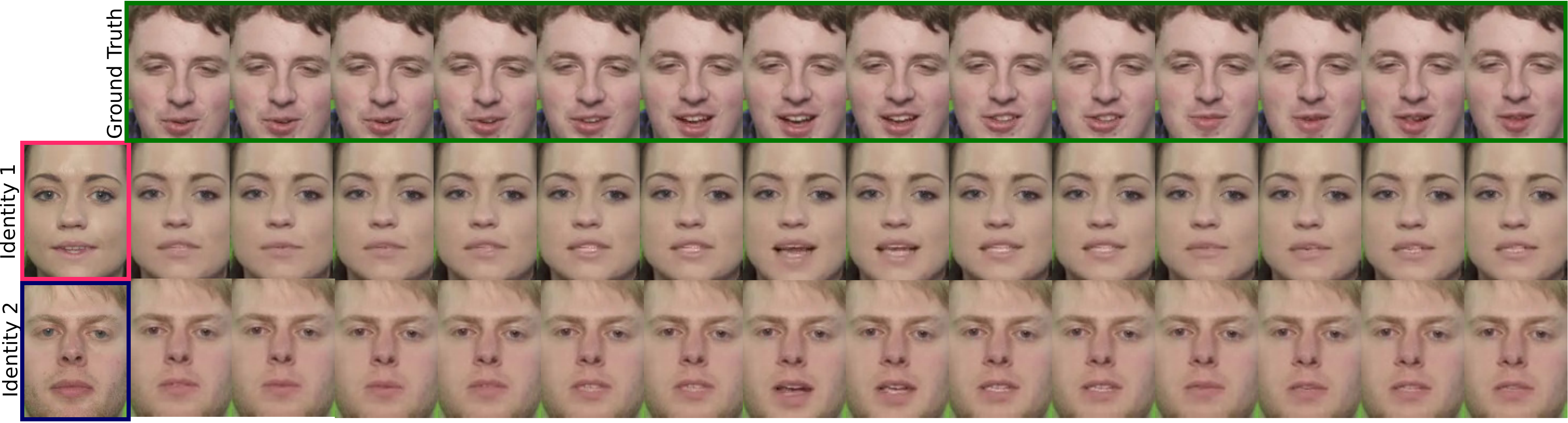}\\
\end{center}
\caption{Animation of different faces using the same audio. The movement of the mouth is similar for both faces as well as for the ground truth sequence. Both audio and still image are unseen during training.}
\label{fig:diffface}
\end{figure}

\begin{figure}[h!]
  \centering
  \begin{subfigure}[b]{0.49\linewidth}
    \centering\includegraphics[width=0.99\textwidth]{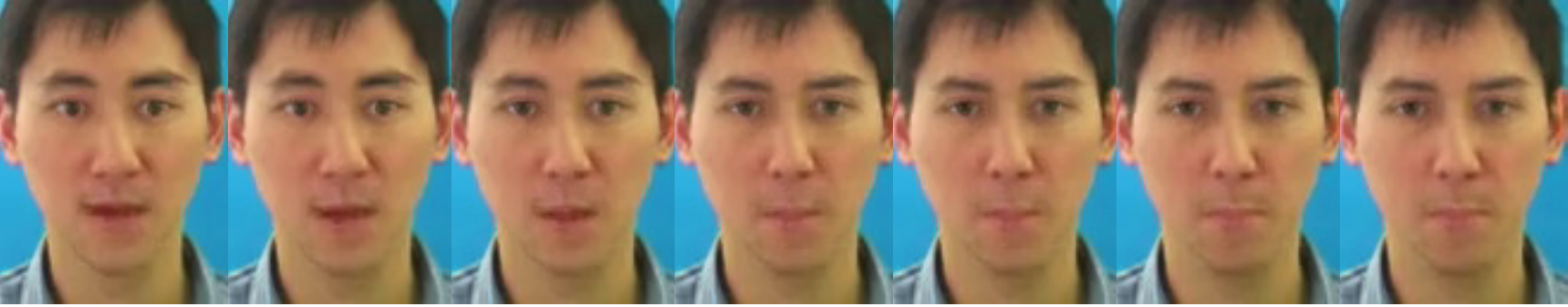}
    \caption{\label{fig:frown}  Example of generated frown}
  \end{subfigure}
  \begin{subfigure}[b]{0.49\linewidth}
    \centering\includegraphics[width=0.99\textwidth]{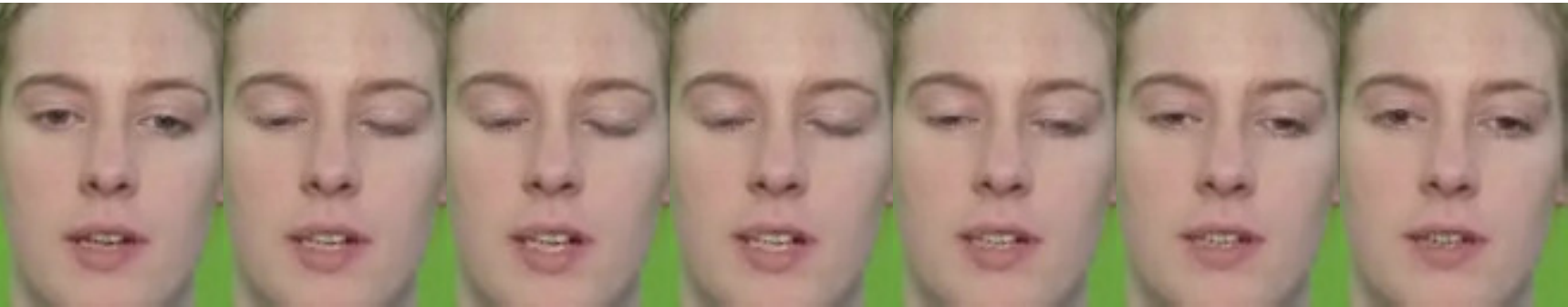}
    \caption{\label{fig:blink}   Example of generated blink}
  \end{subfigure}
  \caption{Facial expressions generated using our framework include (a) frowns and (b) blinks.}
\label{fig:expressions}
\end{figure}

Our method is capable of producing realistic videos of previously unseen faces and audio clips taken from the test set. The examples in \figref{fig:diff_audio} show the same face animated using sentences from different subjects (male and female). The same audio used on different identities is shown in \figref{fig:diffface}. From visual inspection it is evident that the lips are consistently moving similarly to the ground truth video. Our method not only produces accurate lip movements but also natural videos that display characteristic human expressions such as frowns and blinks, examples of which are shown in \figref{fig:expressions}.

\begin{figure}[h!]
  \centering
  \begin{subfigure}[b]{0.49\linewidth}
    \centering\includegraphics[width=0.99\textwidth]{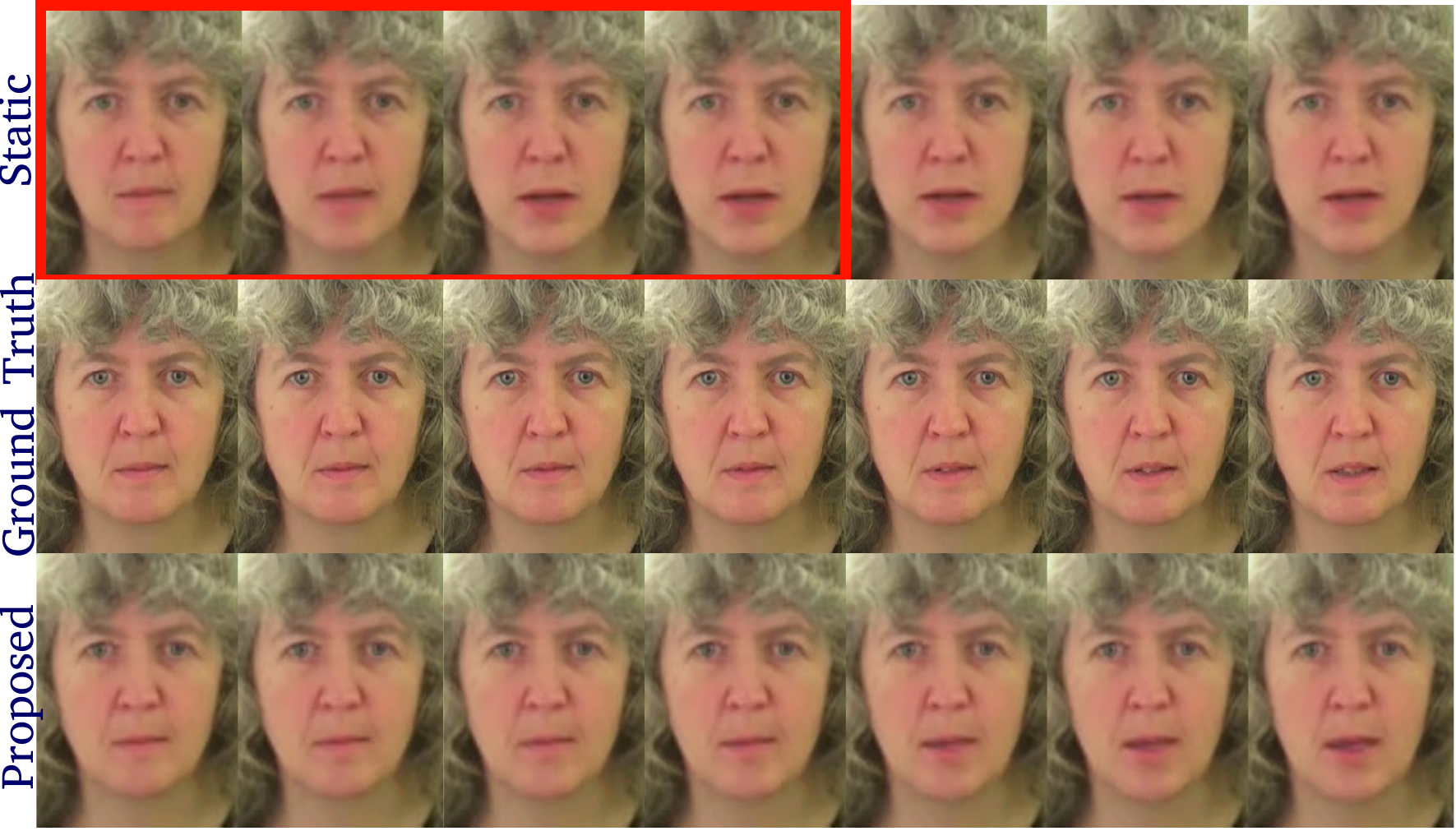}
    \caption{\label{fig:jitterface_silence}  Audio-visual inconsistency during silence}
  \end{subfigure}
  \begin{subfigure}[b]{0.49\linewidth}
    \centering\includegraphics[width=0.99\textwidth]{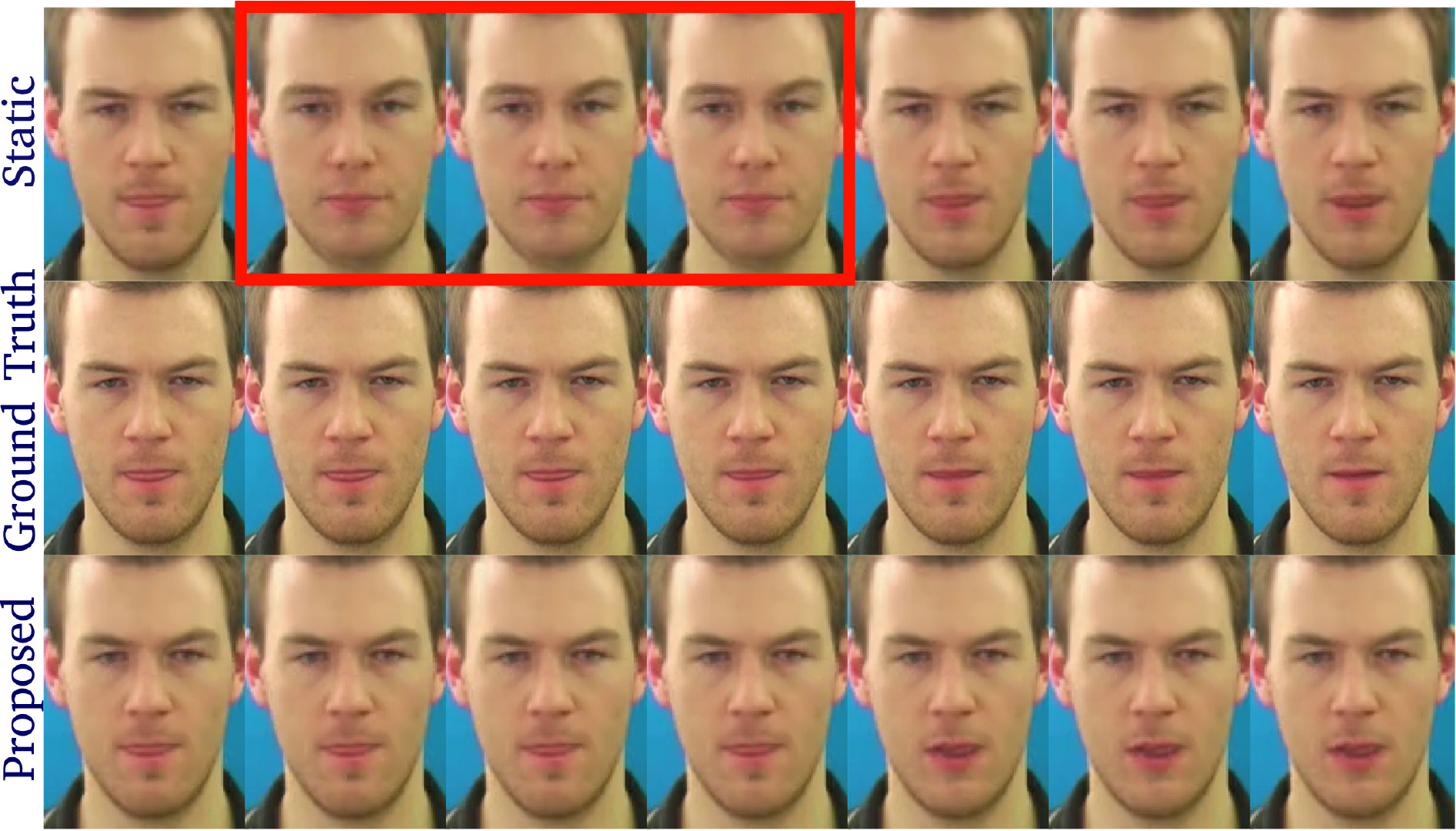}
    \caption{\label{fig:jitterface_cont_break}   Extraneous frames that break continuity}
  \end{subfigure}
  \caption{Examples of consecutive frames showcasing the failures of the static baseline including (a) opening the mouth when words are not spoken (b) producing irrelevant frames that do not take into account the previous face state, thus breaking the sequence continuity.}
\label{fig:jitterface}
\end{figure}

The works that are closest to ours are those proposed in \cite{Suwajanakorn2017} and \cite{Chung2017}. The former method is subject dependent and requires a large amount of data for a specific person to generate videos. For the latter method there is no publicly available implementation so we compare our model to a static method that produces video frames using a sliding window of audio samples like that used in \cite{Chung2017}. This is a GAN-based method that uses a combination of an $L_1$ loss and an adversarial loss on individual frames. We will also use this method as the baseline for our quantitative assessment in the following section. This baseline produces less coherent sequences,  characterized by jitter, which becomes worse in cases where the audio is silent (e.g. pauses between words). This is likely due to the fact that there are multiple mouth shapes that correspond to silence and since the model has no knowledge of its past state generates them at random. \figref{fig:jitterface} shows a comparison between our approach and the baseline in such cases.

\vspace{-5pt}

\subsection{Quantitative Results}
\label{sec:quantitative}
We measure the performance of our model on the GRID and TCD TIMIT datasets using the metrics proposed in section \ref{sec:metrics} and compare it to the static baseline. Additionally, we present the results of a 30-person survey, where users were shown 30 videos from each method and were asked to pick the more natural ones. The results in \tabref{tab:quantitative} show that our method outperforms the static baseline in both frame quality and content accuracy. Although the difference in performance is slight for frame-based measures (e.g. PSNR, ACD) it is substantial in terms of user preference and lipreading WER,  where temporal smoothness of the video and natural expressions play a significant role.

 \begin{table}[h!]
 \centering
 \begin{tabular}{|c|l|r|r|r|r|r|r|r|}
 \hline
 & \multicolumn{1}{c|}{Method}& \multicolumn{1}{c|}{PSNR} & \multicolumn{1}{c|}{SSIM} & \multicolumn{1}{c|}{FDBM} & \multicolumn{1}{c|}{CPBD} & \multicolumn{1}{c|}{ACD}& \multicolumn{1}{c|}{User} & \multicolumn{1}{c|}{WER}\\
 \hline\hline
 \small \parbox[t]{2mm}{\multirow{2}{*}{\rotatebox[origin=c]{90}{GRID}}} 
 & Proposed Model &  \b{27.98} & \b{0.844} & \b{0.114} & 0.277     & \b{1.02} $\cdot 10^{-4}$ & \b{79.77}\% & \b{25.4}\% \\
 & Baseline       &     27.39  & 0.831     & 0.113 & \b{0.280} & 1.07 $\cdot 10^{-4}$ & 20.22\% & 37.2\% \\
 \hline
 \small \parbox[t]{2mm}{\multirow{2}{*}{\rotatebox[origin=c]{90}{TCD }}}
& Proposed Model & \b{23.54} & \b{0.697} & \b{0.102} & \b{0.253} & \b{2.06} $\cdot 10^{-4}$ &\b{77.03\%}&N/A\\
& Baseline       &  23.01 & 0.654     & 0.097 & 0.252 & 2.29 $\cdot 10^{-4}$ & 22.97\%&N/A\\
 \hline
 \end{tabular}
 \caption{Performance comparison of the proposed method against the static baseline. The pretrained LipNet model is not available for the TCD TIMIT so the WER metric is omitted.}
\label{tab:quantitative}
 \end{table}

\begin{figure}[h!]
\begin{center}
\includegraphics[width=0.99\textwidth]{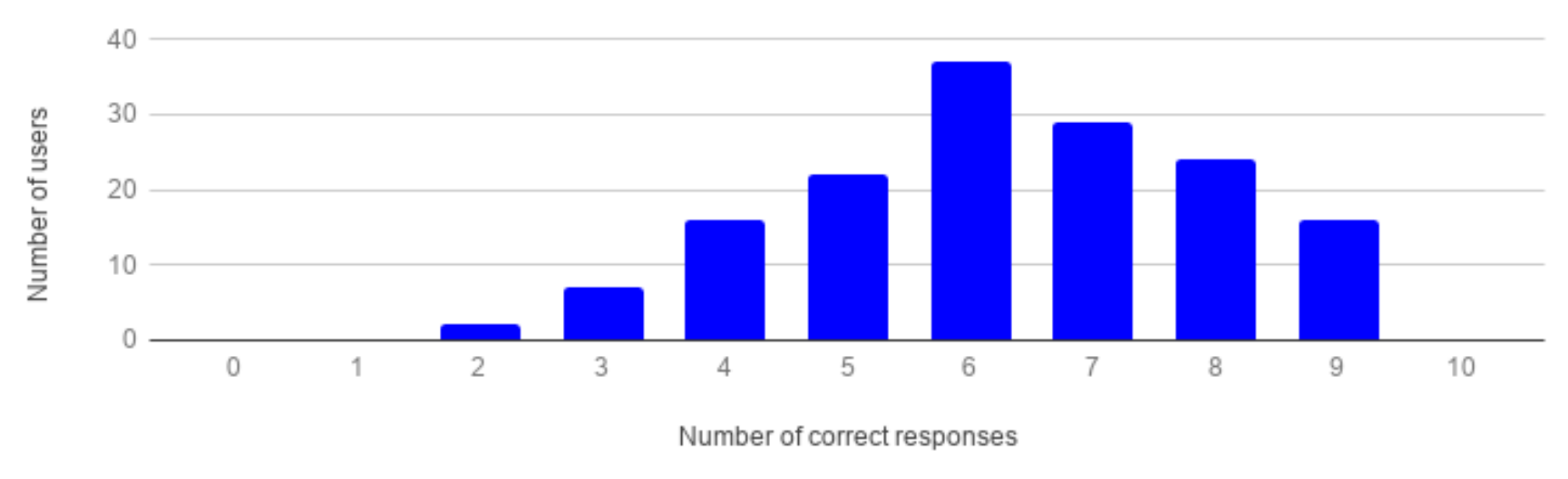}\\
\end{center}
\vspace{-20pt}
\caption{Distribution of user scores for the Turing test.}
\label{fig:response_chart}
\end{figure}

We further evaluate the realism of the generated videos through an online Turing test. In this test users are shown 10 videos, which were chosen at random from GRID and TIMIT consisting of 6 fake videos and 4 real ones. Users are shown the videos in sequence and are asked to label them as real or fake. Responses from 153 users were collected with the average user labeling correctly 63\% of the videos. The distribution of user scores is shown in \figref{fig:response_chart}.

\section{Conclusion and Future Work}
\label{sec:conclusion}
In this work we have presented an end-to-end model using temporal GANs for speech-driven facial animation. Our model is capable of producing highly detailed frames scoring high in terms of PSNR, SSIM and in terms of the sharpness measures on both datasets. According to our ablation study this can be mainly attributed to the use of a {\em Frame Discriminator}.

Furthermore, our method produces more coherent sequences and more accurate mouth movements compared to the static approach, as demonstrated by a resounding user preference and the difference in the WER. We believe that these improvements are not only a result of using a temporal generator but also due to the use of the conditional {\em Sequence Discriminator}. Unlike previous approaches \cite{Chung2017} that prohibit the generation of facial expressions, the adversarial loss on the entire sequence encourages spontaneous facial gestures. This has been demonstrated with examples of blinks and frowns. All of the above factors make the videos generated using our approach difficult to separate from real videos as revealed from the Turing test results, with the average user scoring only slightly better than chance. It is also noteworthy that no user was able to perfectly classify the videos.

This model has shown promising results in generating lifelike videos. Moving forward, we believe that different architectures for the sequence discriminator could help produce more natural sequences. Finally, at the moment expressions are generated randomly by the model so a natural extension of this method would attempt to also capture the mood of the speaker from his voice and reflect it in the facial expressions. 

\section{Acknowledgements}
This work has been funded by the European Community Horizon 2020 under grant agreement
no. 645094 (SEWA).

\appendix
\section{Supplementary Material}
Details regarding the network architecture that were not included in the paper due to lack of space are included here.

\subsection{Audio Preprocessing}
The sequence of audio samples is divided into overlapping audio frames in a way that ensures a one-to-one correspondence with the video frames. In order to achieve this we pad the audio sequence on both ends and use the following formula for the stride:
\begin{equation}
stride=\frac{rate_{audio}}{rate_{video}}
\label{eq:cutting_stride}
\end{equation}

\subsection{Network Architecture}
This section describes, in detail, the architecture of the networks used in our temporal GAN. All our networks use $ReLU$ activations except for the final layers. The encoders and generator use the hyperbolic tangent activation to ensure that their output lies in the set $[-1,1]$ and the discriminator uses a Sigmoid activation.

\subsection{Audio Encoder}
The {\em Audio Encoder} network obtains features for each audio frame. It is made up of 7 Layers and produces an encoding of size $256$. This encoding is fed into a 2 layer GRU which will produce the final context encoding.

\begin{figure}[h!]
\begin{center}
\includegraphics[width=0.7\textwidth]{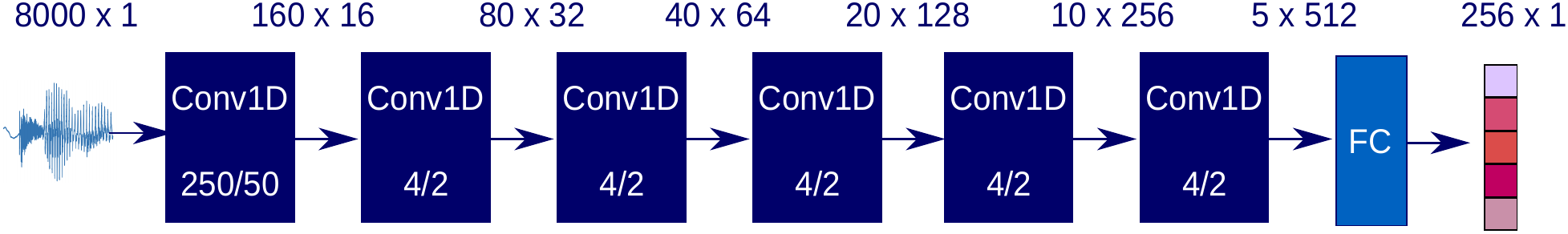}
\end{center}
\caption{The deep audio encoder used to extract 256 dimensional features from audio frames containing 8000 samples. Convolutions are described using the notation {\em kernel / stride}. The feature dimensions after each layer are shown above the network using the notation {\em feature size $\times$ number of feature maps}.}
\label{fig:audio_encoder}
\end{figure}
\vspace{-6pt}

\subsubsection{Noise Generator}
\begin{figure}[h!]
\begin{center}
\includegraphics[width=0.41\textwidth]{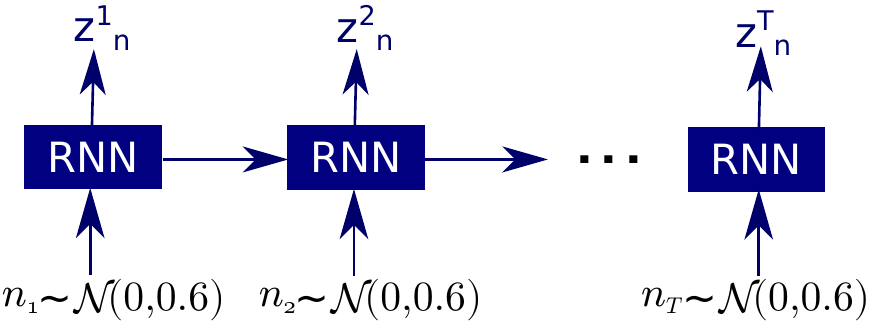}
\end{center}
\vspace{-5pt}
\caption{The network that generates the sequential noise}
\label{fig:noise_gen}
\end{figure}

The {\em Noise Generator} is responsible for producing noise that is sequentially coherent. The network is made up of GRUs which take as input at every instant a 10 dimensional vector sampled from a Gaussian distribution with mean $0$ and variance of $0.6$. The {\em Noise Generator} is shown in \figref{fig:noise_gen}.

\subsubsection{Identity Encoder and Frame Decoder}
\begin{figure}[h!]
\begin{center}
\includegraphics[width=.95\textwidth]{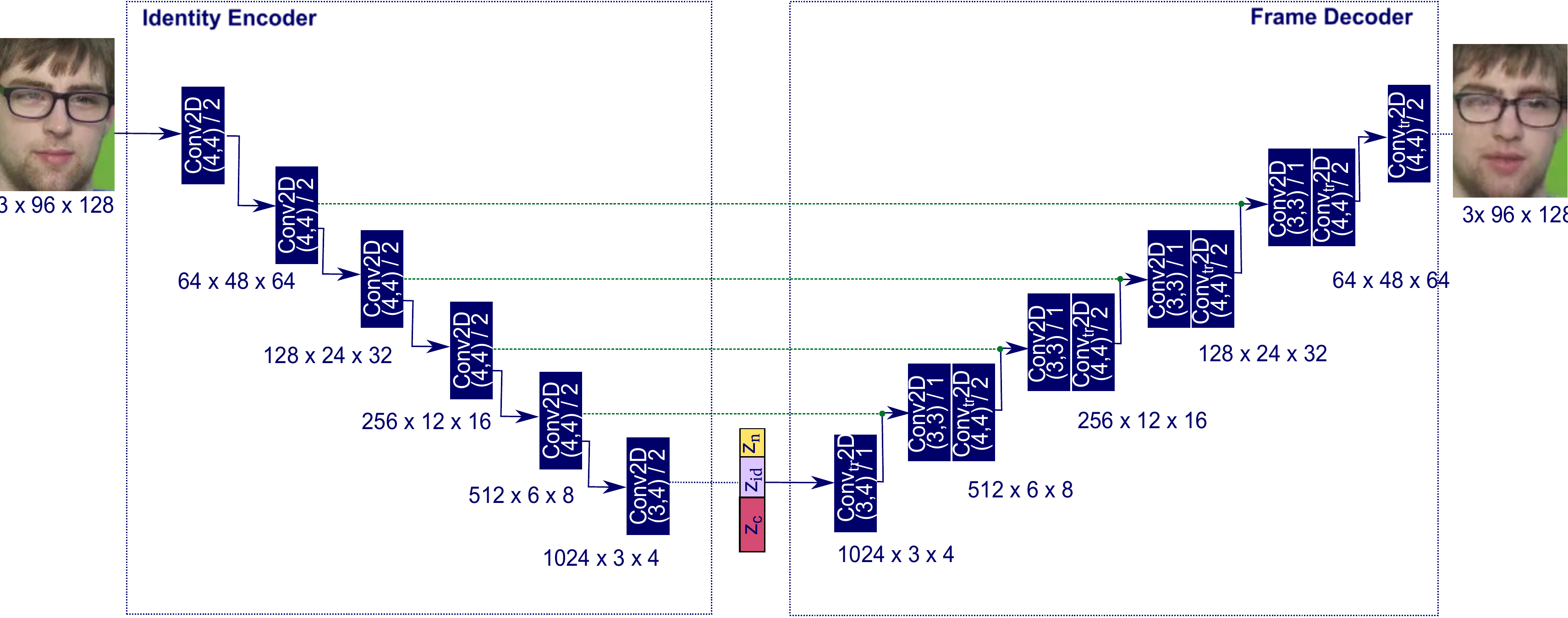}
\end{center}
\caption{The U-Net architecture used in the system with skip connections from the hidden layers of the {\em Identity Encoder} to the {\em Frame Decoder}. Convolutions are denoted by {\em Conv2D} and transpose convolutions as {\em $Conv_{tr}2D$}. We use the notation {\em ($kernel_x$, $kernel_y$) / stride} for 2D convolutional layers.}
\label{fig:enc_dec}
\end{figure}

The {\em Identity Encoder} is responsible for capturing the identity of the speaker from the still image. The {\em Identity Encoder} is a 6 layer CNN which produces an identity encoding $z_{id}$ of size $50$. This information is concatenated to the context encoding $z_c$ and the noise vector $z_n$ at every instant and fed as input to the {\em Frame Decoder}, which will generate a frame of the sequence. The {\em Frame Decoder} is a 6 layer CNN that uses strided transpose convolutions to generate frames. The {\em Identity Encoder} -  {\em Frame Decoder} architecture is shown in \figref{fig:enc_dec}

\subsection{Datasets}
The model is evaluated on the GRID and TCD TIMIT datasets. The subjects used for training, validation and testing are shown in \tabref{tab:train_val_test_subjects}

\begin{table}[h!]
\begin{center}
\begin{tabular}{|l|p{6cm}|p{1.5cm}|p{2cm}|}
\hline
\multicolumn{1}{|c|}{Dataset} & \multicolumn{1}{c|}{Training} & \multicolumn{1}{c|}{Validation}& \multicolumn{1}{c|}{Testing} \\
\hline\hline
GRID & 1, 3, 5, 6, 7, 8, 10, 12, 14, 16, 17, 22, 26, 28, 32 & 9, 20, 23, 27, 29, 30, 34 &  2, 4, 11, 13, 15, 18, 19, 25, 31, 33 \\
\hline
TCD TIMIT         & 1, 2, 3, 4, 5, 6, 7, 10, 11, 12, 13, 14, 16, 17, 19, 20, 21, 22, 23, 24, 26, 27, 29, 30, 31, 32, 35, 37, 38, 39, 40, 42, 43, 46, 47, 48, 50, 51, 52, 53, 57, 59& 34, 36, 44, 45, 49, 54, 58 &8, 9, 15, 18, 25, 28, 33, 41, 55, 56  \\

\hline
\end{tabular}
\end{center}
\caption{The subject IDs for the training, validation and test sets for the GRID and TCD TIMIT dataset.}
\label{tab:train_val_test_subjects}
\end{table}

\bibliography{Mendeley_Speech2Vid.bib}
\end{document}